# Fano Resonance in a Subwavelength Mie-based Metamolecule with Split Ring Resonator

Xiaobo Wang, Ji Zhou[a]

State Key Laboratory of New Ceramics and Fine Processing, School of Materials Science and Engineering, Tsinghua University, Beijing 100084, P. R. China

*Abstract*

In this letter, we report a method of symmetry-breaking in an artificial Mie-based metamolecule. A Fano resonance with a Q factor of 96 is observed at microwave frequencies in a structure combining a split ring resonator (SRR) and a high-permittivity dielectric cube. Calculations indicate resonant frequency tunability will result from altering the cube's permittivity. The asymmetric spectrum is attributed to both constructive and destructive near-field interactions between the two distinct resonators. Experimental data and simulation results are in good agreement. The underlying physics is seen in field distribution and dipole analysis. This work substantiates an approach for the manipulation of Mie resonances which can potentially be utilized in light modulating and sensing.

Metamaterials are artificial structures whose constituent units (metamolecules) have strong electromagnetic properties and are subwavelength in size. They have been the subject of much investigation over the past decade. They allow significant freedom of design, leading to rather intriguing physical properties such as negative refractive index[1,2], and light transformation[3]. Based on various resonators, such as split ring resonators (SRRs) or variants, electromagnetic metamaterials are able to function as invisible cloaks[4,5], and perfect lenses[2]. The Fano resonance, a quantum phenomenon exhibiting a non-Lorentzian spectral shape[6,7] and closely related to electromagnetically induced transparency (EIT), can be obtained by a symmetry-breaking method[8-12] in metamaterials. In this way, a dark mode is able to interfere with the bright resonant one[7,12], leading to both destructive and constructive coupling in the near field and an asymmetric lineshape. Fano resonances normally produce a sharp spectrum, with a high Q-factor and an extremely confined field within the structure[13], indicating a potential for sensing and near-field mapping applications[14-16]. Moreover, a nonreciprocal Fano resonance with tunability has also been seen in previous research[10], suitable for one-way applications.

---

[a] Author to whom correspondence should be addressed. Electronic mail: zhouji@mail.tsinghua.edu.cn

Several metamaterial structures, such as asymmetric split ring resonators (ASRs)[9], mirror-symmetric joint split ring resonators (JSRRs)[17] and dielectric-based resonators[18,19], have shown the Fano response under various interaction mechanisms. Subwavelength metallic and dielectric resonators are typical building blocks in the construction of electromagnetic metamaterials. Metallic components will dissipate electromagnetic energy and become quite lossy at high frequencies[20], while fully dielectric-based metamaterials with low heat losses overcome such obstacles. A so-called Mie resonance[20-22] based on Mie scattering theory is often used as the active element in dielectric-based metamaterials. However, one may consider combining metallic and dielectric components to form a hybrid metamolecule. Gu[23] *et al*. reported an actively reconfigurable EIT metamaterial in the terahertz region utilizing cut wires and SRRs with silicon islands. Ke Bi[24] *et al*. demonstrated enhanced magnetic resonance in a Mie metamaterial with the inclusion of SRRs. Despite similar insight in previous research, not much work has been focused on Fano interference inside such composites. In this letter, we have designed a ring-shaped Mie-based metamolecule displaying an asymmetric resonance at microwave frequencies and the Fano response was observed in experiments. Symmetry breaking was achieved by introducing a metallic resonator. The metamolecule consists of an SRR with a high-permittivity ceramic cube located at the gap. The Mie resonance appears inside the dielectric cube while the SRR adds an LC circuit response. According to our simulation, the asymmetric Fano profile results from interference between dipoles with different physical natures.

The schematic diagram of the metamolecule is shown in Fig. 1(a). A copper SRR is clad on a low-index Rogers RO4350B PCB board ($\varepsilon = 3.66 + 0.0037i$) and a dielectric cube penetrates the board at the SRR gap. The copper is 0.035 mm thick and the board is 0.762 mm thick. The outer dimension of the assembled subwavelength metamolecule is $10 \times 10$ mm, approximately 1/3 of the working wavelength. The relative permittivity of the cube is $154 + 0.0024i$ with an edge length of 2 mm, exactly matching the gap. The inner and the outer radii of the SRR are 2 mm and 3 mm. The incident EM plane wave is polarized along the *y*-direction, perpendicular to the gap in the model. A simulation of the interaction between the metamolecule and X-band microwave radiation was performed in the commercial finite-element solver Comsol Multiphysics.

We began by both simulating and experimentally measuring absorption spectra for the dielectric cube alone. Then we repeated both on the complete metamolecule assembly including the copper SRR. For further understanding of the Fano phenomenon, the field and current distributions at resonance were simulated. We then calculated the electric and magnetic dipole moment distributions in both the cube and the SRR. Finally, we focused on the tunability of the metamolecule by altering the permittivity of the cube. Details of the experimental methods are summarized at the end of this paper[25].

Figures 2(a,b) give the calculated and experimental transmission responses $S_{21}$ of the dielectric cube and the metamolecule for normal incidence. Fig. 2(b) indicates that the assembly exhibits a Fano-type resonance at 10.13 GHz, which is adjacent to the first-order Mie resonance frequency of the dielectric cube. The single dielectric cube shows a symmetric Lorentz lineshape, suggesting the introduction of metallic resonator alters the transmission spectrum of the cube. The experimentally measured metamolecule transmission dip is located at 10.14 GHz with a low point of 0.132 or -17.6 dB, and the surrounding maximum level is 0.943 or -0.51 dB. These agree well with our numerical simulation in both cases. The measured 3 dB bandwidth of the metamolecule is 0.106 GHz. The calculated Q factor of the resonance is 96. In Figs. 2(c,d) the magnetic field along the *x* axis is enhanced and the electric field circulates along the *x* axis inside the dielectric cube, denoting a strong magnetic dipole or $TE_{011}$ mode. The planar distribution of electric field and the induced current around the SRR in Fig. 2(d) are characteristic of a dark mode. We notice there is current flow near the cube-SRR interface region because of strong localized electric field density. The simulation shows a much wider feature, a broad-band SRR resonance with a transmission dip located at the much higher frequency 19.91 GHz. Summarizing simply, the dielectric block presents a typical first-order

Mie resonance, while the copper ring shows an LC circuit response.

We further analyze the electric and magnetic dipole moment components along the three axes inside each resonator and the radiated power of the whole structure in the far field. The calculations utilize the volume integrals[8]

$$p(\omega) = -\frac{1}{i\omega}\iiint J(\omega,r)\mathrm{d}V. \quad (1)$$

$$m(\omega) = \frac{1}{2}\iiint r \times J(\omega,r)\mathrm{d}V. \quad (2)$$

$$W_p(\omega) = \frac{k_0^4}{12\pi\eta_0\varepsilon_0^2}|p(\omega)|^2. \quad (3)$$

$$W_m(\omega) = \frac{k_0^4\eta_0}{12\pi}|m(\omega)|^2. \quad (4)$$

where $k_0$ is the wave number, $\varepsilon_0$ and $\eta_0$ are the permittivity and impedance of free space, and $J(r)$ is the current density at the point $r = (x,y,z)$ including displacement and induced currents. We rewrite the formulas in component form during the calculation. In Figs. 3(a-c), the magnetic dipole radiates strongly near the Fano resonance frequency and the magnetic dipole moment along the $x$ direction is the largest component inside the dielectric cube. Meanwhile, the electric dipole generates a weak radiation background in far field corresponding to the maximum component along $z$ axis inside the SRR. The quadrupoles are not presented in this letter as they are found to be negligible in the calculation. This is consistent with our earlier analysis based on the field distribution. The $S_{21}$ phase variation with frequency in both the individual components and the metamolecule assembly is depicted in Fig. 3(d). Both destructive and constructive coupling bracket the transmission dip. Near field interference between two modes from different physical mechanisms causes the Fano resonance and the confined field. A similar discovery was found at optical frequencies involving nanoscale assemblies[26], in that introducing a semi-infinite dielectric substrate produced plasmonic mode hybridization of a metallic nanostructure.

Introducing a metallic SRR changes the transmission spectrum of the dielectric cube, which interacts with microwaves through Mie resonance. Hence, the metamolecule assembly allows two methods of tuning the Fano resonance, one through the SRR and the second through the dielectric cube. As an example, from Lewin's theory[22,27], we derive the first Mie resonance frequency for a lossless dielectric sphere of radius $r$ and relative permittivity $\varepsilon$

$$f = c\theta/2\pi r\sqrt{\varepsilon}. \qquad (5)$$

where $c$ is the speed of light in vacuum, and $\theta$ is a constant roughly equivalent to $\pi$. Then we have

$$f \cong c/2r\sqrt{\varepsilon}. \qquad (6)$$

We are thus able to tune the Fano resonance by altering the Mie parameters, the dimensions and permittivity of the cube. Figure 4 shows simulated transmission spectra in parametric sweep, illustrating how permittivity influences the microwave absorption by the structure. As permittivity increases from 110 to 180, the resonant frequency of the isolated cube is red shifted from 11.94 to 9.38 GHz. The Fano absorption resonance in the metamolecule behaves nearly the same way, although at a few discrete permittivity values, the resonance is shifted slightly. The resonance frequencies roughly satisfy the $f \propto 1/\sqrt{\varepsilon}$ relation. Unlike the isolated cube Mie resonance profile, the Fano dip is asymmetric, broader on the blue side of resonance, implying a lower figure of merit. The broad variation of resonant frequency with cube permittivity verifies that our design is capable of good tunability over a certain frequency range.

$CaTiO_3$ ceramic doped with 1 wt. % $ZrO_2$ was selected as the dielectric cube material for its suitable dielectric constant. Commercially available $CaTiO_3$ and $ZrO_2$ powders were utilized as raw materials. The mixed powders were ball-milled and after dehydration and granulation were pressed into flakes by cold isostatic pressing at 200 MPa. Flakes were sintered at 1350 °C for 2 h in a Nabertherm muffle furnace. The resulting ceramic plates were cut into cubes using a dicing saw. The SRRs were fabricated via laser processing of commercially available copper clad laminates. The square hole on the substrate was punched by machine tool. Finally, the dielectric cubes were press fit into the holes to complete the assembly. The S parameter measurements were carried out in an X-band BJ-100 rectangular waveguide ($22.86 \times 10.16$ mm) by an Agilent N5230C Network Analyzer. As only the $TE_{01}$ mode can propagate in the waveguide, samples were placed at the center of the cross section using double sided tape.

In conclusion, we have demonstrated Fano resonance in a subwavelength metamolecule interacting with X-band microwaves. The metamolecule is constructed from a high-permittivity ceramic cube which undergoes Mie resonance, coupled to a copper SRR. We have both modeled the response numerically and measured transmission spectra in the 9 to 12 GHz region. The Q factor of the resonance reaches 96 at 10.14 GHz. There is a good agreement between simulation and experiment. Computational results demonstrate that current and electric field assembling exist at contact points between the two resonators. Our dipole and field distribution analysis indicate a strong magnetic mode and weak electric mode emerge. Fano resonance tunability is predicted via simulation. As the permittivity of dielectric cube increases from 110 to 180, the transmission dip moves down from 11.9 to 9.38 GHz. This work offers a method for modulation of Mie-based metamaterials which can be applied to metadevice sensors.

This work was supported by the National Natural Science Foundation of China under grants 11274198 and 51532004.